\documentclass[fleqn,usenatbib]{mnras}

\usepackage{newtxtext,newtxmath}

\usepackage[T1]{fontenc}
\usepackage{ae,aecompl}

\usepackage{graphicx}	
\usepackage{amsmath}	

\usepackage{amssymb}	
\usepackage{lipsum}
\usepackage{pdflscape}
\usepackage{afterpage}
\usepackage{multirow}

\title[Mrk~335]{A systematic study of photoionized emission and warm absorption signatures of the NLS1 Mrk 335}

\author[H. Liu et al.]{H. Liu,$^{1}$\thanks{E-mail: hhliu19@fudan.edu.cn}
M. L. Parker,$^{2,3}$
J. Jiang,$^{4}$
E. Kara,$^{5}$
Cosimo Bambi,$^{1}$
D. Grupe, $^{6}$
and S. Komossa$^{7}$
\\
$^{1}$Center for Field Theory and Particle Physics and Department of Physics, Fudan University, 200438 Shanghai, China\\
$^{2}$European Space Agency (ESA), European Space Astronomy Centre (ESAC), E-28691 Villanueva de la Ca\~{n}ada, Madrid, Spain\\
$^{3}$Institute of Astronomy, University of Cambridge, Madingley Road, Cambridge CB3 0HA, UK\\
$^{4}$Department of Astronomy, Tsinghua Univerisity, Shuangqing Road, Beijing 100084, China\\
$^{5}$MIT Kavli Institute for Astrophysics and Space Research, MIT, 70 Vassar Street, Cambridge, MA 02139, USA\\
$^{6}$Department of Physics, Earth Science, and Space System Engineering, Morehead State University, 235 Martindale Dr, Morehead, KY40351, USA\\
$^{7}$Max-Planck-Institut f\"ur Radioastronomie, Auf dem H\"ugel 69, 53121 Bonn, Germany
}

\date{Accepted XXX. Received YYY; in original form ZZZ}

\pubyear{2020}

\begin{document}
\label{firstpage}
\pagerange{\pageref{firstpage}--\pageref{lastpage}}
\maketitle

\begin{abstract}
We present an analysis of all the archival high resolution spectra of the Narrow-line Seyfert 1 galaxy Mrk~335 obtained with Reflection Grating Spectrometer (RGS) on board \textit{XMM-Newton}. The spectra show rich emission and absorption features in low and intermediate flux intervals. We model the emission lines with the \textsc{pion\_xs} grid and try to find any possible correlation between the properties of the emitting gas and the source flux. Current data does not allow detailed trace of the response of the line emitting gas to the X-ray flux of Mrk~335, but the flux of the X-ray lines is significantly less variable than the X-ray continuum. We also find that the warm absorber's properties are not correlated with the flux variability. From the latest \textit{XMM-Newton} observation in 2019 December, we find that the photoionized emission and distant reflection components have not responded to the flux drop of Mrk~335 from 2018 July. However, the possible existence of partial covering absorber in the 2018--2019 low state of Mrk~335 makes it difficult to constrain the scale of the emitting gas using this lack of response.
\end{abstract}




\section{Introduction}

Active Galactic Nuclei (AGNs) are powered by matter accretion onto the central supermassive black hole (SMBH), emitting a large amount of energy in the form of radiation. The strong radiation causes ionization of gas in the nuclear environment of AGNs. The ionization state of the photoionized gas can be quantified by the ionization parameter: $$\xi=L/nr^2$$ where $L$ is the ionizing luminosity, $n$ is the gas density and $r$ is the distance between the gas and the X-ray source. In the soft X-ray band, emission lines from the photoionized gas have been observed in both type 1~\citep[e.g.][]{Kaspi2002, Mao2018} and type 2 objects~\citep[e.g.][]{Sambruna2001, Bianchi2006, Marinucci2011}. The emission features are more evident when the observed primary continuum is weak, either being obscured~\citep[e.g.][]{Mao2019} or due to an intrinsic flux drop~\citep[e.g.][]{Peretz2019}. Modeling these emission lines can help to constrain the scale and geometry of the X-ray broad line region and narrow line region~\citep[e.g.][]{Kinkhabwala2002, Longinotti2008, Whewell2015, Grafton-Waters2020}.

Warm absorbers (WA) are the photoionized gas on the line of sight, leaving signatures in the X-ray spectra in the form of blue shifted absorption lines. These absorption lines are commonly seen in the soft X-ray spectra of AGNs~\citep{Reynolds1997, Porquet2004, Laha2014} and can be used to diagnose the ionization state and dynamics of the gas. Although it has been around 30 years since the first observational evidence of the X-ray warm absorbers~\citep{Halpern1984}, their launching mechanism and origin are still not well understood. A one-to-one correspondence between objects that show intrinsic UV absorption and warm X-ray absorption is observed~\citep{Crenshaw1999}, indicating a connection between the two phenomena. There is also some evidence suggesting that WA winds can fit into a large scale outflow scenario~\citep{Tombesi2013}, in which the ultra-fast outflows~\citep[UFOs; see][]{Tombesi2010, Tombesi2012, Parker2017} are launched from the inner disk and WAs from the outer disk or the torus~\citep{Blustin2005}. It is also not clear how the outflows interact with the host galaxy. In some moderate-luminosity AGNs, the outflowing UV and warm X-ray absorption components seem have the potential to drive significant feedback to their host galaxies~\citep{Crenshaw2012}.

One of the key parameters to answer these questions is the location of the warm absorber. Photoionization modelling can measure the column density ($n_{\rm H}$), ionization state ($\xi$) and outflowing velocity ($v$) of the warm absorbers. With the ionizing luminosity ($L$) determined from spectral energy distribution (SED) of the source, we can only obtain the product $nr^2$ using the definition of ionization parameter. To determine $r$, the density $n$ must be known. One of the methods to estimate the density is to measure the response (or lack of response) of the gas to the variations of the ionizing luminosity. The response yields the recombination timescale that depends on the density. This approach has been applied to estimate density of warm absorbers in several well-know variable AGNs, e.g., NGC~3783~\citep{Krongold2005}, NGC~4051~\citep{Krongold2007} Mrk~509~\citep{Kaastra2012} and NGC~5548~\citep{Ebrero2016}, etc. This kind of analysis needs variability of the source and multi-epoch studies of the warm absorbers with high resolution instruments. It can also be challenging if the observations of a given source is sparse. In the case of NGC~5548, \cite{Ebrero2016} was able to test the variability of the 6 WAs for timescales from 2 days to a decade and put stringent constraint on their locations. For NGC~7469, however, only a large upper limit can be obtained for some WA components due to the sparse observations~\citep{Mehdipour2018}.

Mrk~335 ($z=0.025785$) is a Narrow Line Seyfert 1 (NLS1) galaxy with a SMBH of mass $(2.6 \pm 0.8)\times 10^7 M_\odot$~\citep{Grier2012}. The source is strongly variable on both long and short time scales in X-rays~\citep{Grupe2012}. Its X-ray flux has been detected, several times, to change by more than an order of magnitude over the last decade \citep[see][]{Gallo2018}. Moreover, clear X-ray emission lines~\citep[e.g.][]{Longinotti2008, Parker2019} and warm absorption features~\citep{Longinotti2013} have been found in its soft X-ray spectra. Therefore, Mrk~335 is potentially an interesting source to study the response of the line emitting gas and warm absorbers to the variable ionizing continuum. \cite{Longinotti2013} already tested the variability of WAs in Mrk~335 with RGS spectra before 2009. We are now able to investigate the variation on a longer timescale with more observations after 2009 (See Table~\ref{obs}). The other motivation of this work is that different codes have been used to model the WAs (e.g., XSTAR in~\cite{Parker2019}; PHASE in~\cite{Longinotti2013}) in Mrk~335 and we know that the differences between these codes may not be negligible~\citep[see][]{Mehdipour2016}. It is thus useful to re-analyze the absorption and emission signatures of Mrk~335 in all archival spectra with one consistent code.


\section{Observations and data reduction}

Mrk~335 has been observed by \textit{XMM-Newton}~\citep{Jansen2001} ten times, the details of all these observations are shown in Table~\ref{obs}. In this paper, we include all available high-resolution spectra of the Reflection Grating Spectrometer (RGS,~\citet{RGS}) to study the warm absorption and photoionized emission features.

\textit{Swift} started to monitor Mrk~335 in 2007 when its X-ray flux dropped dramatically~\citep{Grupe2007, Grupe2008}. The source was caught entering the lowest UV flux state ever seen by \textit{Swift} in 2018 July. The X-ray flux remained in a low state for about 2 years and was found to rise in May 2020 when Mrk~335 was out of the \textit{Swift} Sun constraint~\citep{Komossa2020}. Two ToO observations on 2018 July 11 and 2019 January 08 by \textit{XMM-Newton} were triggered and analyzed~\citep{Parker2019}. Here we also present the analysis of the third ToO observation on 2019 December 27 (PI Kara) by \textit{XMM-Newton}. 

The data are reduced with the Science Analysis Software version 18.0.0. For the EPIC pn data from 2019 December, we produce calibrated event files with the task \textsc{epproc} and filter the background flaring by setting the threshold of the 10--12 keV light curve. We extract source counts from a circular region (20 arcsec) surrounding the source, including patterns 0--4. Background counts are from a circular region (40 arcsec) near the source. The redistribution matrices and auxiliary response matrices are computed with the tasks \textsc{rmfgen} and \textsc{arfgen}, respectively. The EPIC pn spectrum is then binned to a minimal signal to noise (S/N) of 6 and to oversample the response by a factor of 3.

The data from RGS1 and RGS2 are processed with the \textsc{rgsproc} tool. For each dataset in Table~\ref{obs}, we stack the spectra from RGS1 and RGS2 using the \textsc{rgscombine} tool. We exclude data from the last 20 ks of the 2015 observation because of the flux increase intrinsic to the source (see Figure 1 in~\citet{Longinotti2019}). All RGS spectra are binned by a factor of 8.

\begin{table}
    \centering
    \caption{\textit{XMM} observations of Mrk~335} \label{obs}
    \begin{tabular}{cccccc}
    \hline\hline
    ObsID & Date & Exp (ks) & Flux$^{\rm a}$ & Flux state$^{\rm b}$ & Ref \\
    \hline
    0101040101 & 2000-12-25 & 36.9 & $3.26\pm0.01$ & HS &  1,2\\
    0101040701 & 2000-12-25 & 10.9 & $3.26\pm0.01$ & HS & 1,2\\
    0306870101 & 2006-01-03 & 133.3 & $4.18\pm0.02$ & HS & 3\\
    0510010701 & 2007-07-10 & 22.6 & $0.48\pm0.02$ & LS & 4,5\\
    0600540601 & 2009-06-11 & 132.2 & $0.72\pm0.02$ & MS & 6\\
    0600540501 & 2009-06-13 & 82.6 & $0.91\pm0.01$ & MS & 6\\
    0741280201 & 2015-12-30 & 140.4 & $0.36\pm0.01$ & LS & 7\\
    0780500301 & 2018-07-11 & 114.5 & $0.18\pm0.01$ & LS & 8\\
    0831790601 & 2019-01-08 & 117.8 & $0.17\pm0.01$ & LS & 8\\
    0854590401 & 2019-12-27 & 105.9 & $0.14\pm0.01$ & LS & 9\\
    \hline
    \end{tabular}  

    \textbf{Notes.} (a) X-ray fluxes in 0.5--10 keV (in units of 10$^{-11}$ erg cm$^{-2}$ s$^{-1}$) by EPIC pn data. (b) Flux state for each observation, HS for high state, MS for mid state and LS for low state. For references: (1)~\citet{Gondoin2002}; (2)~\citet{Longinotti2007}; (3)~\citet{ONeill2007}; (4)~\citet{Grupe2008}; (5)~\citet{Longinotti2008}; (6)~\citet{Longinotti2013}; (7)~\citet{Longinotti2019}; (8)~\citet{Parker2019}; (9) Present work.
\end{table}

\section{Spectral analysis}
\label{analysis}

\begin{figure}
    \centering
    \includegraphics[width=\linewidth]{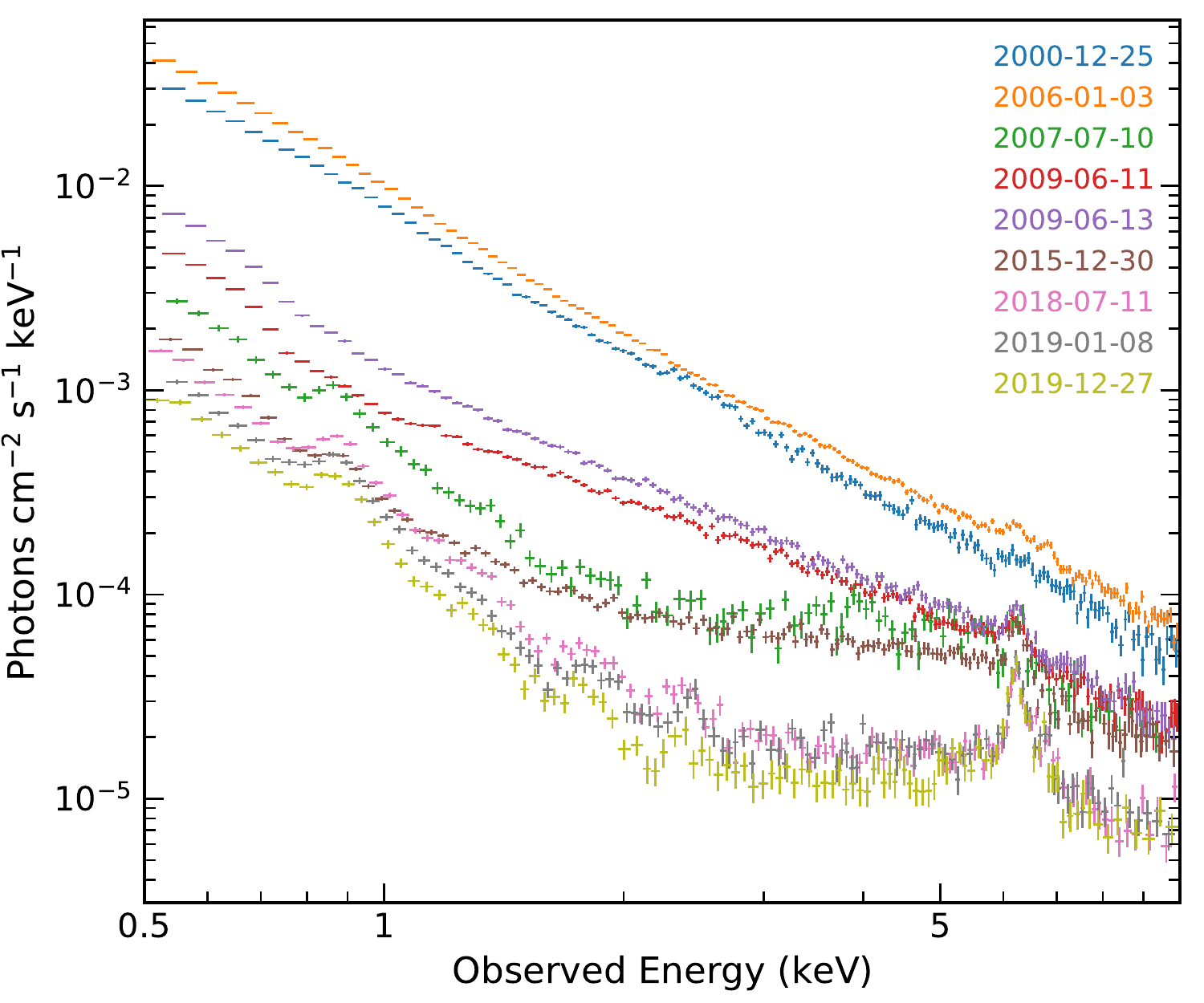}
    \caption{Unfolded spectra of Mrk 335 from the EPIC pn detector. The response is unfolded with \textsc{xspec} assuming a power-law model with zero index.}
    \label{pnspec}
\end{figure}

\begin{figure*}
    \centering
    \includegraphics[width=\linewidth]{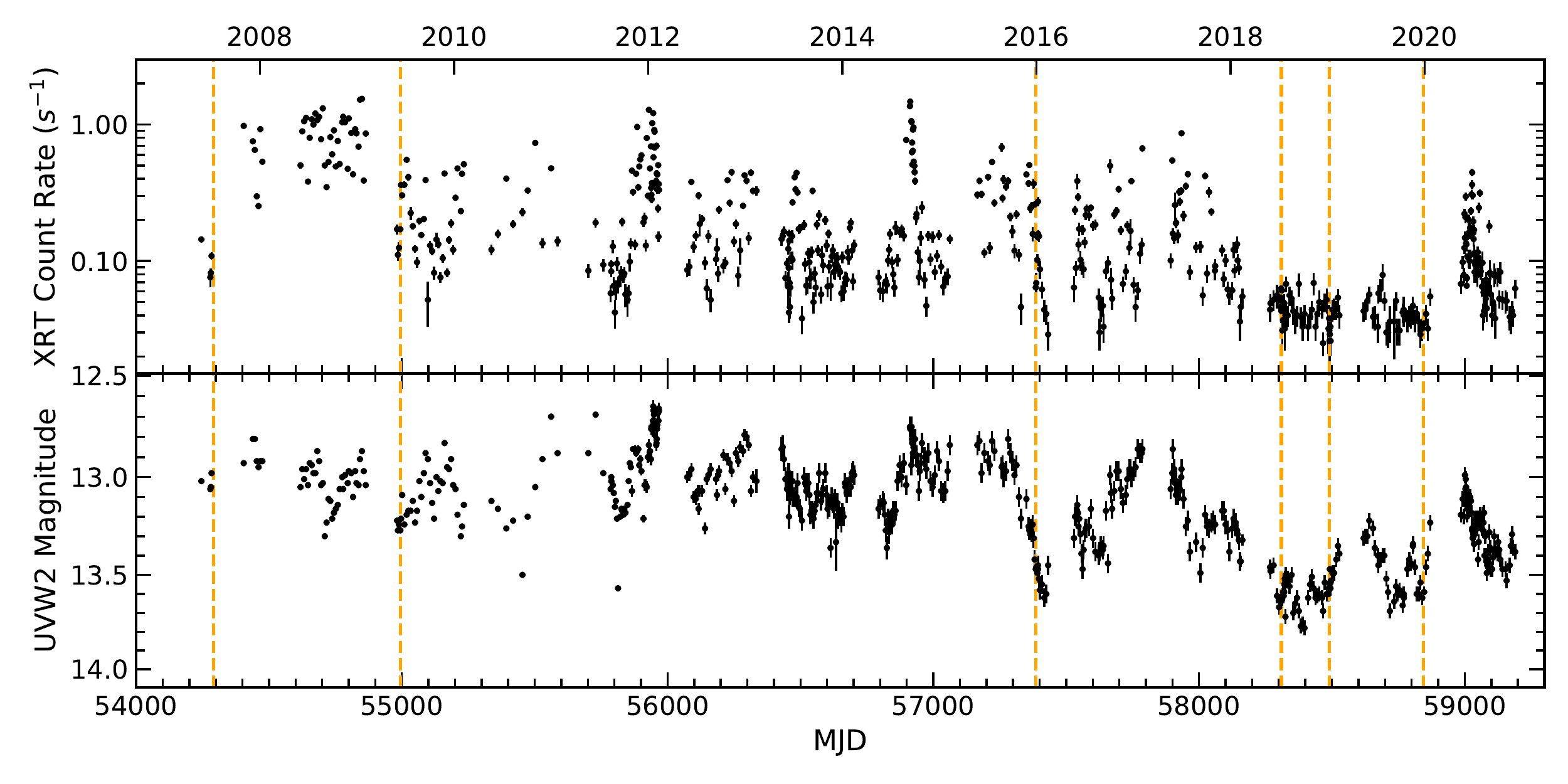}
    \caption{\textit{Swift} monitoring of Mrk~335 since 2007 May. The vertical orange lines mark observations by \textit{XMM-Newton}.}
    \label{lightcurve}
\end{figure*}

This work is mainly intended to study the high resolution RGS spectra of Mrk~335 and to reveal the behavior of warm absorbers and photoionized material of the system. To illustrate the spectral variability of the source, we show all available EPIC pn spectra in Fig.~\ref{pnspec} and the \textit{Swift} monitoring of Mrk~335 since 2007 in Fig.~\ref{lightcurve}. Note that the UVW2 light curve shown in Fig.~\ref{lightcurve} is processed using the new UVOT calibration (20200925). The old calibration caused an over-correction of the UV data since 2017 which gradually increased to about 0.35 mag in W2 by the end of 2020.\footnote{\url{https://www.swift.ac.uk/analysis/uvot/index.php}} We divide all the RGS spectra into  high-state (2000, 2006), mid-state (2009) and low state as done by~\citet{Longinotti2013}.

The analysis in this paper is conducted with \textsc{xspec} v12.10.1f \citep{xspec}. We use solar element abundances from~\citet{Lodders2009} and cross sections from~\citet{Verner1996}. We calculate a grid of table models with \textsc{xstar}~\citep{Kallman2001} for each observation to model possible warm absorption features. The absorption medium in \textsc{xstar} is assumed to have a spherical geometry with constant density. We also assume a turbulent velocity of 200 km s$^{-1}$ and the solar abundances~\citep{Grevesse1996}. The input spectral energy distribution (SED) to \textsc{xstar} is obtained by fitting the RGS + pn data with a continuum model: \textsc{tbnew$\times$(powerlaw + bbody)} and extrapolating the best-fit to the UV end, with the absorption corrected. To fit photoionized emission lines, we use \textsc{pion\_xs}\footnote{\url{https://bit.ly/pion_xs}}~\citep{Parker2019}, an \textsc{xspec} table model calculated from the \textsc{pion}~\citep{Miller15_pion} model of \textsc{spex}~\citep{spex}. The \textsc{pion\_xs} table model needs less computation power than its analytical version. A slab geometry of the gas is assumed when calculating the table model. It also assumes a power-law input spectrum with $\Gamma=2$, a velocity broadening of 100 km s$^{-1}$ and the abundances of \cite{Lodders2009}.There might be concerns for using different abundances for the emission and absorption material, but we note that the abundances for important elements (e.g. C, O, N and Fe) are consistent within the uncertainties (below 30\%) between \cite{Grevesse1996} and \cite{Lodders2009}. \footnote{We have also tested, in the absorption dominated spectrum (2009, See Sec.~\ref{analysis:2009}), that the differences caused by abundance versions are minor compared to the statistical uncertainties for the parameters of \textsc{xstar} (e.g., the ionization parameter of the highest ionization component changes from $3.18_{-0.13}^{+0.2}$ to $3.1_{-0.5}^{+0.3}$ if using abundances of~\citet{Lodders2009}.).} We also include the model \textsc{redge} for radiative recombination continua (RRC) signatures. A Galactic absorption with $N_{\rm H}=3.56\times 10^{20}$ cm$^{-2}$~\citep{Kalberla2005} is always added to fit the spectra. Cash statistics~\citep{Cash1979} is applied to the analysis of RGS spectra.

The baseline model we use for RGS spectral analysis is: \textsc{tbnew$\times$ xstar$\times$ (powerlaw + pion\_xs + redge)}. We show only one layer of \textsc{xstar} and \textsc{pion\_xs} for clarity. Multiple layers of these components may be needed for the fitting (see details of the analysis below).

\subsection{The 2000 and 2006 high-state spectra}

Two spectra from 2000 December 25 are combined using tool \textsc{rgscombine} before the fitting. In this high state, the emission lines are buried in the strong primary continuum. We find that one layer of warm absorption with a shallow column density is able to fit the spectra well. The best-fit warm absorption parameters are given in Table~\ref{RGS:WA}. A single layer of \textsc{pion\_xs} is enough to model the emission lines in 2000 spectrum. Adding a second layer does not improve the fit statistics much ($\Delta$ C-stat of 9 with 4 degrees of freedom). However, for the 2006 data, two \textsc{pion\_xs} components are needed to fully describe all lines. The best-fit values for the emission material are listed in Table~\ref{RGS:PE}. The lower ionization component mainly accounts for \ion{O}{vii} and \ion{C}{vi} lines. The higher ionization component is mainly fitting the \ion{O}{viii} line.

\subsection{The 2009 mid-state spectrum}
\label{analysis:2009}

We combine RGS spectra from the 2 data sets in 2009 since there is no significant variation~\citep{Longinotti2013}. The 2009 spectrum exhibits the most complex absorption and emission signatures. The photon index of the power-law component is hard to constrain because of the narrow energy coverage of RGS and the complexity of the model. We thus fix it to the value found by simultaneously fitting RGS + pn spectrum and include a \textsc{diskbb} component to fit the soft excess. We find three layers of warm absorber (see Table~\ref{RGS:WA}) as reported by~\citet{Longinotti2013}. Although the outflow velocities relative to the source are free to vary, they end up sharing a consistent value ($\sim$ 5200 km s$^{-1}$). This consistency in velocity strongly suggests that the three absorbers are part of the same complex wind, which presumably has a clumpy structure. The ionization parameters, defined as $\xi=L/nr^2$, span from $\log(\xi)=1.32$ to $\log(\xi)=3.18$. The absorber with a higher column density is associated with a higher ionization parameter. 

To fit the emission features, we include three \textsc{pion\_xs} components (see Table~\ref{RGS:PE}). Using one layer of \textsc{pion\_xs} would leave significant residuals for \ion{O}{vii} and \ion{O}{viii} emission lines. The second and third layers improve the C-stat by 40 and 30 respectively with 4 degrees of freedom. Though having the same number of layers as that of warm absorbers, the emitting clouds show very different patterns. The ionization states of the three phases of emitting material are quite similar. Significant distinctions between velocities of the emitting phases are detected, which is quite different from the consistency of velocities of the warm absorbers. This discrepancy in velocity structure may suggest that the emitters (at least the two with lower velocities) have a different origin from that of the absorbers.

\subsection{The low-state spectra}
\label{analysis:low}

Since the flux contribution from the continuum is low, the low-state spectra reveal rich features of emission lines. Nonetheless, also due to the low continuum flux, the absorption features are hard to detect. One layer of warm absorber is enough to fit the absorption for all low-state spectra.

We find two photoionized emitters and two RRC signatures of \ion{O}{viii} and \ion{C}{vi} in 2007 spectrum. The low ionization layer improves the C-stat by 15 and is necessary to fit the \ion{O}{vii} He$\alpha$ lines. The ionization states and velocities of the two emitters are close to that found in 2006 high-state spectrum. This consistency is interesting since it suggests the emitters are not responding to the flux drop in 2007. Using the same data, \citet{Longinotti2008} find no evidence for line-of-sight absorption. We test the existence of warm absorption by including a \textsc{xstar} component. Indeed, we find this component is not significantly required since the improvement on the fit is minor ($\Delta$ C-stat $=9$) and the absorption column is relatively shallow ($\sim 10^{20}$ cm$^{-2}$).

There is a clear absorption signature around 16~\AA\ from the Fe unresolved transition array~\citep{Netzer2004} in the 2015 RGS spectrum (see Figure 5 of \citet{Longinotti2019}). Fitting the absorption with \textsc{xstar} gives the highest column density in low-state spectra. The properties we find about this absorber are consistent with \citet{Longinotti2019}. For the emission lines, one \textsc{pion\_xs} component already gives an acceptable fit. Including a second layer improves the C-stat by only 7 and the ionization parameter is not well constrained.

As shown in Fig.~\ref{pnspec}, the spectral shape of Mrk~335 above 5~keV has been quite stable since 2018 July. As for the high resolution RGS spectra, there are changes in the line ratios of O triplet. The resonance line dominates in 2018 July, while the 2019 Jan spectrum is dominated by the forbidden line. \citet{Parker2019} fit the two spectra simultaneously and suggest that the variations of emission lines is not due to changes of emitting gas properties, but is a consequence of variations of warm absorption. For the new \textit{XMM} observation in 2019 December, we first fit the RGS spectrum individually and find similar parameters of the absorber and emitters as found by \citet{Parker2019}. This similarity suggests that we are seeing the same material and it is natural to fit the three spectra simultaneously. We include one \textsc{xstar} and two \textsc{pion\_xs} components for each observation to model the absorption and emission features. A \textsc{redge} component is also needed for the \ion{C}{vi} RRC (25.3~\AA~at rest frame). The outflow velocity of the absorber, the velocity and number density of emitters are linked across observations during the fitting. Best-fit parameters are listed in Table~\ref{RGS:WA} and Table~\ref{RGS:PE}. The column density of the warm absorption component is continuously increasing, though with large uncertainties. The photoionized emission consists of a lower and a higher ionization component. The difference in the ionization parameter between observations is minor compared with uncertainties.

\begin{table*}
    \renewcommand\arraystretch{1.5}
    \centering
    \caption{Best-fit parameters for warm absorbers from RGS spectra}
    \label{RGS:WA}
    \begin{tabular}{cccccccc}
    \hline\hline
    Obs & Flux state & WA phase & $\log(\xi)$ & $N_{\rm H}$ (cm$^{-2}$) & $v_{\rm out}$ (km s$^{-1}$) & $\Gamma$  & C-stat/dof \\
    \hline
    2000 & HS & \romannumeral1 & $1.2_{-0.5}^{+0.2}$    & $7_{-3}^{+5} \times 10^{19}$       & $-2500_{-350}^{+300}$ & $2.834_{-0.015}^{+0.015}$  & 423/378\\
    \hline
    2006 & HS & \romannumeral1 & $1.23_{-0.18}^{+0.3}$ & $2.6_{-1.4}^{+2} \times 10^{19}$ & $-7500_{-500}^{+600}$ & $2.705_{-0.009}^{+0.010}$  & 430/375\\
    \hline
    2007 & LS& \romannumeral1 & $0.7_{-0.4}^{+0.4}$    & $2.0_{-0.9}^{+1.9} \times 10^{20}$ & $-6500_{-500}^{+200}$ & $2.68_{-0.14}^{+0.12}$  & 361/357\\
    \hline
    2009 & MS & \romannumeral1 & $1.32_{-0.07}^{+0.03}$ & $2.1_{-0.3}^{+0.3} \times 10^{21}$ & $-5200_{-70}^{+70}$   & $1.57^*$  & 572/362 \\
         && \romannumeral2 & $2.35_{-0.07}^{+0.07}$ & $5.8_{-1.5}^{+1.6} \times 10^{21}$ & $-5200_{-90}^{+120}$   & &\\
         && \romannumeral3 & $3.18_{-0.13}^{+0.2}$ & $2.1_{-1.6}^{+5} \times 10^{22}$ & $-5200_{-450}^{+400}$  & &\\
    \hline
    2015 & LS & \romannumeral1 & $1.53_{-0.09}^{+0.08}$ & $3.5_{-1.1}^{+1.3} \times 10^{21}$ & $-6000_{-150}^{+120}$ & $2.62_{-0.15}^{+0.15}$ & 439/371 \\
    \hline
    2018 & LS & \romannumeral1 & $1.36_{-0.13}^{+0.10}$ & $2.9_{-1.0}^{+1.3} \times 10^{20}$ &  \multirow{3}{*}{$-5700_{-200}^{+100}$}  & \multirow{3}{*}{$2.19_{-0.18}^{+0.17}$} & \multirow{3}{*}{1176/924}\\
    \cline{1-4}
    2019 Jan & LS & \romannumeral1 & $1.26_{-0.11}^{+0.10}$ & $5_{-2}^{+2} \times 10^{20}$ &   & & \\
    \cline{1-4}
    2019 Dec & LS & \romannumeral1 & $1.36_{-0.08}^{+0.08}$ & $7_{-2}^{+4} \times 10^{20}$ &   & & \\
    \hline
    \end{tabular}

    \textit{Note.} The parameter $v_{\rm out}$ is defined to be negative for outflows. The outflow velocity is linked when simultaneously fitting the 3 spectra from 2018 to 2019. The power-law index of 2009 spectrum is fixed at the value found by fitting RGS + pn data for reasons discussed in Sec.~\ref{analysis:2009}. The last column shows the best-fit statistics when both absorption and emission are considered.
\end{table*}

\begin{table}
    \renewcommand\arraystretch{1.5}
    \centering
    \caption{Best-fit parameters for photoionized emission material from RGS spectra}
    \label{RGS:PE}
    \resizebox{0.49\textwidth}{4.5cm}{
    \begin{tabular}{ccccccc}
    \hline\hline
    Obs & Flux state & phase & $\log(\xi)$  & $v_{\rm out}$ (km s$^{-1}$) & Density (cm$^{-3}$) & Flux$^{\dag}$ \\
    \hline
    2000 & HS & \romannumeral1 & $1.2_{-0.4}^{+0.2}$    & $-2750_{-300}^{+250}$     & $>2.4\times 10^{11}$ & $5_{-2}^{+2}$ \\
    \hline
    2006 & HS & \romannumeral1 & $0.15_{-0.14}^{+0.18}$   & $-3600_{-200}^{+300}$     & $<1.0\times 10^{12}$ & $2.0_{-0.5}^{+0.5}$ \\
         &  & \romannumeral2 & $1.3_{-0.2}^{+0.4}$   & $-370_{-280}^{+260}$      & $<1.0\times 10^{12}$ & $1.3_{-0.4}^{+0.5}$ \\
    \hline
    2007 & LS & \romannumeral1 & $0.26_{-0.14}^{+0.18}$   & $-3300_{-300}^{+300}$     & $<1.0\times 10^{12}$ & $1.9_{-0.5}^{+0.5}$ \\
         &  & \romannumeral2 & $1.40_{-0.14}^{+0.14}$   & $-740_{-740}^{+260}$      & $>2.6\times 10^{11}$ & $2.0_{-0.5}^{+0.6}$ \\
    \hline
    2009 & MS & \romannumeral1 & $1.34_{-0.13}^{+0.17}$   & $-5900_{-300}^{+100}$     & $<4.9\times 10^{11}$ & $1.6_{-0.7}^{+0.9}$ \\
         &  & \romannumeral2 & $1.30_{-0.11}^{+0.11}$   & $-850_{-200}^{+120}$      & $<5.0\times 10^{11}$ & $1.0_{-0.3}^{+0.2}$ \\
         &  & \romannumeral3 & $1.19_{-0.16}^{+0.14}$   & $400_{-280}^{+230}$       & $<1.0\times 10^{12}$ & $1.4_{-0.3}^{+0.4}$ \\
    \hline
    2015 & LS & \romannumeral1 & $1.25_{-0.08}^{+0.08}$   & $-280_{-90}^{+120}$       & $<1.1\times 10^{11}$ & $2.6_{-0.5}^{+0.6}$ \\
    \hline
    2018 & LS  &\romannumeral1 & $0.99_{-0.10}^{+0.08}$   & $-150_{-110}^{+80}$ $^*$      & $<1.3\times 10^{11}$ $^*$ & $1.9_{-0.3}^{+0.3}$ \\
         &  & \romannumeral2 & $2.28_{-0.07}^{+0.08}$   & $-200_{-90}^{+100}$ $^*$      & $<1.0\times 10^{12}$ $^*$ & $1.4_{-0.2}^{+0.2}$ \\
    \hline
    2019 Jan & LS & \romannumeral1 & $0.90_{-0.12}^{+0.11}$ & $-150_{-110}^{+80}$ $^*$       & $<1.3\times 10^{11}$ $^*$ & $1.8_{-0.3}^{+0.3}$  \\
             &  & \romannumeral2 & $2.36_{-0.06}^{+0.06}$ & $-200_{-90}^{+100}$ $^*$     & $<1.0\times 10^{12}$ $^*$ & $1.5_{-0.3}^{+0.3}$  \\
    \hline
    2019 Dec & LS & \romannumeral1 & $1.02_{-0.11}^{+0.13}$ & $-150_{-110}^{+80}$ $^*$      & $<1.3\times 10^{11}$ $^*$ & $1.5_{-0.3}^{+0.3}$ \\
             &  & \romannumeral2 & $2.18_{-0.15}^{+0.11}$ & $-200_{-90}^{+100}$ $^*$    & $<1.0\times 10^{12}$ $^*$ & $1.4_{-0.5}^{+0.3}$ \\
    \hline
    \end{tabular}
    }
    \textit{Note.} The photoionized emission lines are modeled with \textsc{pion\_xs} component. The velocity and density of the emitting material of 2018 and 2019 spectra (marked with $^*$) are tied during the fitting. $^{\dag}$ Flux of the \textsc{pion} components (in units of 10$^{-13}$ erg cm$^{-2}$ s$^{-1}$) in energy range 0.5--2.0 keV calculated with the \textsc{xspec} model \textsc{cflux}.
\end{table}

\begin{figure*}
    \centering
    \includegraphics[width=\linewidth]{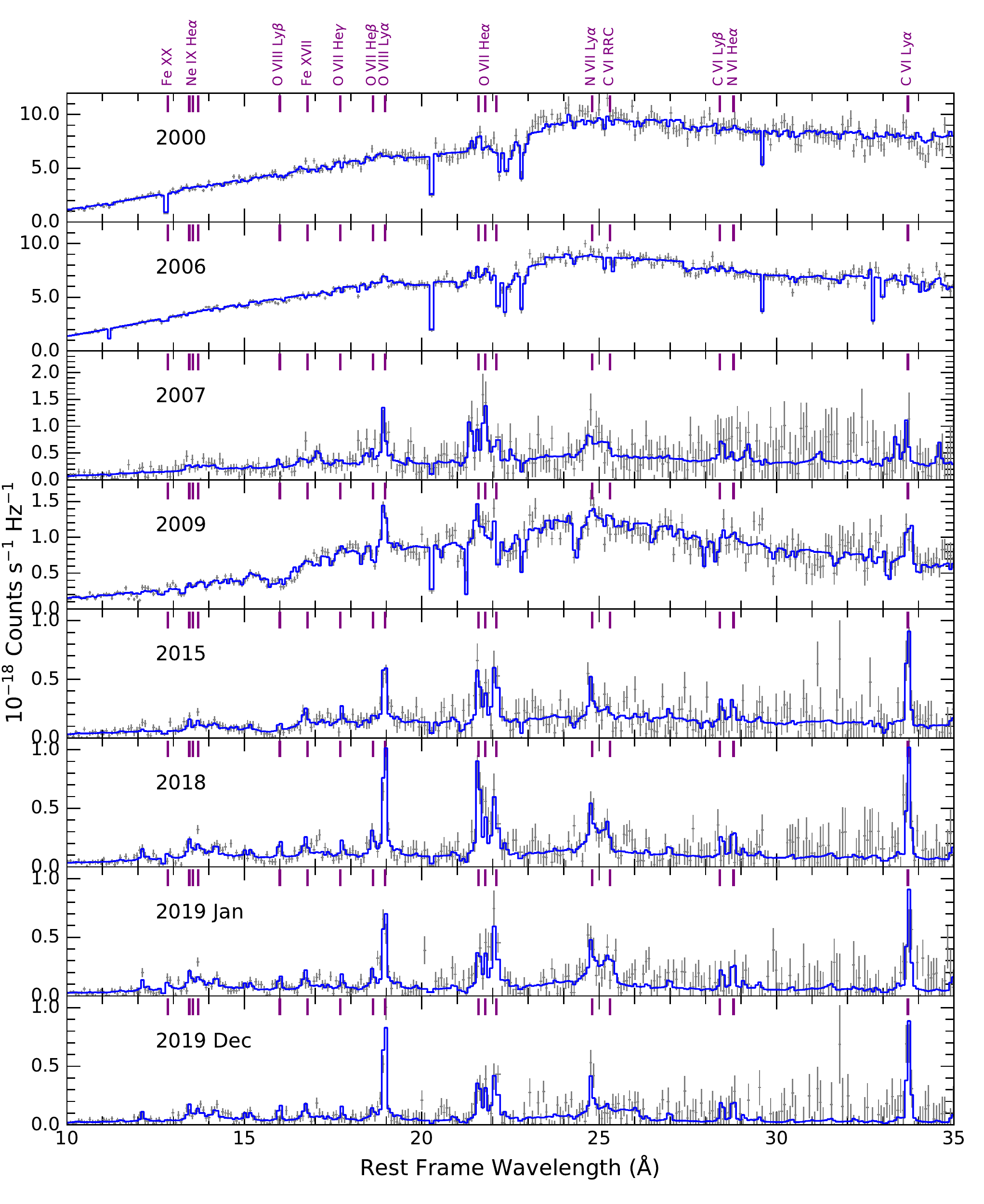}
    \caption{Best-fit models for all RGS spectra analyzed in this work. Emission lines are clearly detected in low-states and are modelled with the component \textsc{pion\_xs}. The warm absorption features are also taken into account with \textsc{xstar}. Details of the analysis can be found in Sec.~\ref{analysis}.}
    \label{RGS:models}
\end{figure*}

\subsection{Broad band fitting}
\label{fitting:broad}
The UV flux of Mrk~335 was observed to decline to the lowest value ever in 2018 July by \textit{Swift}, while its broad band spectrum was still dominated by photoionized emission and distant reflection 6 months after the flux drop~\citep{Parker2019}. This lack of response helps to put a lower limit on the location of the line emitting region and the distant reflector. It would be interesting to see if these components are still dominating the spectrum in 2019 December, which is 17 months after the source entering the low flux state.

\begin{figure}
    \centering
    \includegraphics[width=\linewidth]{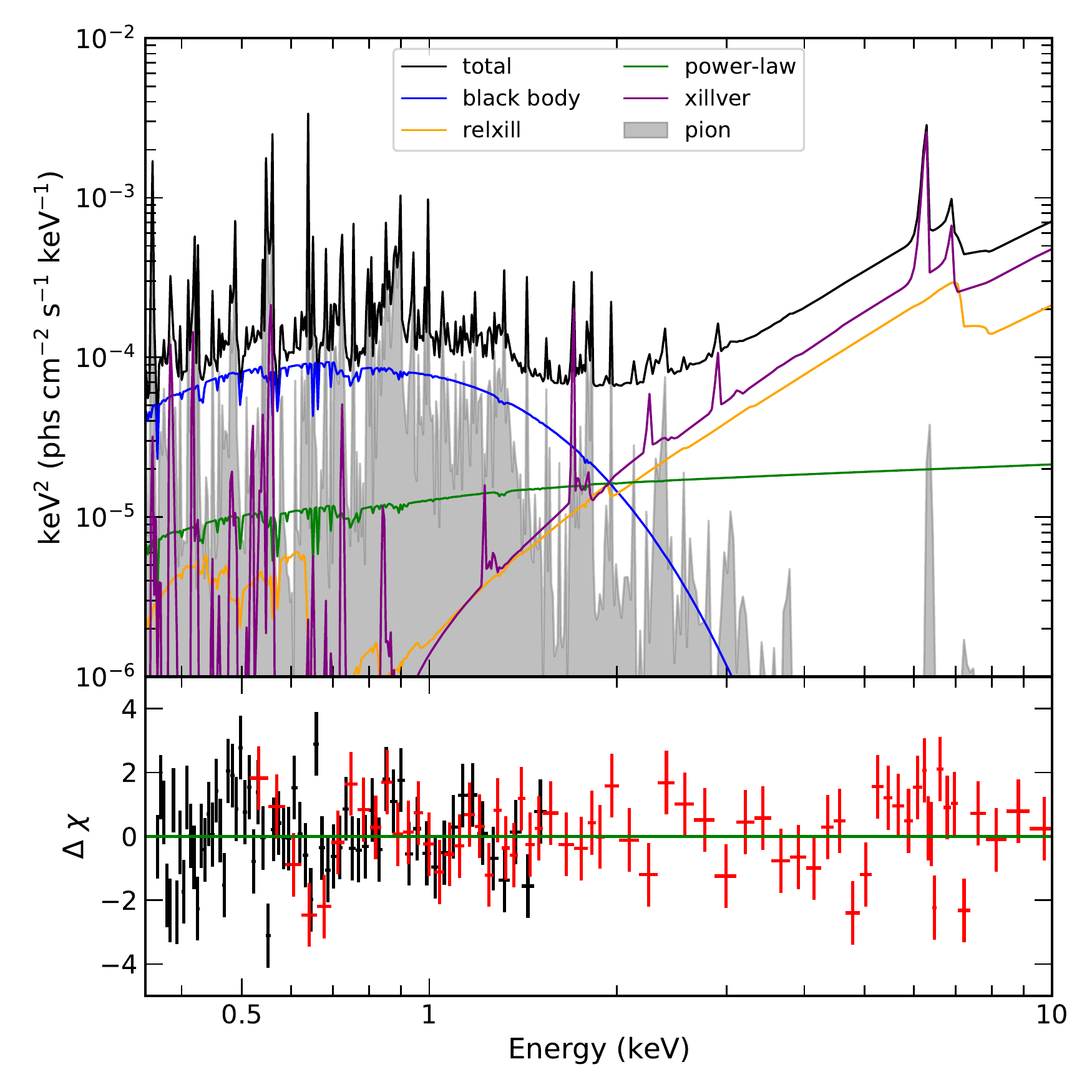}
    \caption{Best-fit model components (top) and residuals (bottom) for the RGS+pn spectra of Mrk~335 from 2019 Dec. Data is fitted with relativistic reflection, distant reflection, continuum and some absorption and emission components.}
    \label{broad:models}
\end{figure}

We fit the RGS and pn spectra from 2018 and 2019 simultaneously. The model used, expressed in \textsc{xspec} notation, is: \textsc{tbnew $\times$ (xstar $\times$ (zpowerlw + diskbb + relxill + pion\_xs + pion\_xs) + xillver)}. The disk blackbody component is used to account for the soft excess that is present in all spectra. Note that there are mainly two scenarios to explain the origin of this component, e.g., warm corona emission~\citep{Matt2014, Petrucci2018} and blurred reflection~\citep{Crummy2006, Garcia2016, Jiang2019}. A detailed study of these scenarios is out of the scope of this paper. The \textsc{relxill}~\citep{Garcia2014} component is included to fit possible relativistic reflection from the accretion disk. The best-fit parameters are shown in Table~\ref{fits:broad}. The best-fit model components and residuals for the 2019 December spectrum are plotted in Fig.~\ref{broad:models}. The plot reveals that the 2019 December spectrum of Mrk~335 is still dominated by photoionized emission and distant reflection.

\begin{table*}
    \centering
    \caption{Best-fit values for the joint fit to RGS + pn data with relativistic reflection model.}
    \label{fits:broad}
    \renewcommand\arraystretch{1.5}
    \begin{tabular}{lccccc}
        \hline\hline
        Description & Component & Parameter & 2018 & 2019 Jan & 2019 Dec \\
        \hline
        Galactic absorption & \textsc{tbnew} & $N_{\rm H}$ (10$^{20}$ ${\rm cm}^{-2})$ & $3.56^*$ & $3.56^*$ & $3.56^*$ \\
        \hline
        Warm absorption & \textsc{xstar} & $N_{\rm H}$ (10$^{21}$ ${\rm cm}^{-2})$ & $1.26_{-0.4}^{+0.4}$ & $1.0_{-0.3}^{+0.4}$ & $1.0_{-0.2}^{+0.3}$  \\
        & & $\log(\xi)$ & $1.44_{-0.06}^{+0.07}$ & $1.25_{-0.18}^{+0.11}$ & $1.34_{-0.11}^{+0.06}$  \\
        & & $v$ (km s$^{-1}$) & \multicolumn{3}{c}{$-6100_{-70}^{+60}$} \\
        \hline
        Soft excess & \textsc{diskbb} & $T_{\rm in}$ (keV) & $0.249_{-0.007}^{+0.016}$ & $0.217_{-0.017}^{+0.014}$& $0.27_{-0.02}^{+0.02}$  \\
        &  & Norm & $7_{-2}^{+3}$ & $10_{-3}^{+5}$ & $2.6_{-1.0}^{+1.3}$  \\
        \hline
        Continuum & \textsc{zpowerlw} & $\Gamma$ & $1.97_{-0.10}^{+0.06}$ & $1.85_{-0.3}^{+0.07}$ & $1.85_{-0.12}^{+0.07}$  \\
        &  & Norm ($10^{-5}$) & $5.8_{-0.6}^{+0.7}$ & $4.6_{-0.3}^{+0.6}$ & $1.3_{-0.5}^{+0.8}$  \\
        \hline
        Emission (hot) & \textsc{pion\_xs} & $\log(\xi)$ & $2.38_{-0.03}^{+0.03}$ & $2.36_{-0.04}^{+0.04}$ & $2.31_{-0.04}^{+0.04}$  \\
        & & $v$ (km s$^{-1}$) & \multicolumn{3}{c}{$-200_{-60}^{+60}$} \\
        & & Density (cm$^{-3}$) &  \multicolumn{3}{c}{$<1\times 10^{12}$} \\
        & & Flux (10$^{-13}$ erg s$^{-1}$ cm$^{-2}$) & $1.9_{-0.2}^{+0.2}$  & $1.6_{-0.2}^{+0.2}$ & $1.4_{-0.2}^{+0.2}$ \\
        \hline
        Emission (cold) & \textsc{pion\_xs} & $\log(\xi)$ & $0.90_{-0.09}^{+0.08}$  & $0.88_{-0.06}^{+0.11}$ & $0.96_{-0.11}^{+0.07}$ \\
        & & $v$ (km s$^{-1}$) & \multicolumn{3}{c}{$-300_{-40}^{+40}$} \\
        & & Density (cm$^{-3}$) &  \multicolumn{3}{c}{$<2.1\times 10^{10}$} \\
        & & Flux (10$^{-13}$ erg s$^{-1}$ cm$^{-2}$) & $2.02_{-0.14}^{+0.3}$ & $1.5_{-0.2}^{+0.3}$ & $1.5_{-0.2}^{+0.2}$  \\
        \hline
        Cold reflection & \textsc{xillver} & $E_{\rm cut}$ (keV) & \multicolumn{3}{c}{$300^*$} \\
        &  & Norm ($10^{-5}$) & $2.8_{-0.3}^{+0.3}$ & $3.2_{-0.4}^{+0.4}$ & $3.2_{-0.4}^{+0.3}$ \\
        \hline
        Ionized reflection & \textsc{relxill} & $q$ & \multicolumn{3}{c}{$3^*$} \\
        &  & $a_*$ & \multicolumn{3}{c}{$0.98^*$} \\
        &  & $i$ (deg) & \multicolumn{3}{c}{$50^*$}\\
        &  & $\log\xi$ & \multicolumn{3}{c}{$<0.05$}\\
        &  & $A_{\rm Fe}$ & \multicolumn{3}{c}{$2.7_{-0.6}^{+0.6}$} \\
        &  & Norm ($10^{-6}$) & $7.7_{-1.2}^{+1.0}$ & $7.3_{-1.4}^{+1.5}$ & $5.2_{-1.2}^{+1.2}$  \\
        \hline
        &  $\chi^2/\nu$ & & \multicolumn{3}{c}{1795/1327} \\
        \hline
    \end{tabular}

    \textit{Note.} Parameters with $^*$ are not well constrained with current spectra and are fixed at given values in the table during the fitting as in~\citet{Parker2019}. Only one value is given for parameters that are tied between the 3 spectra. Flux of the \textsc{pion} components is calculated in 0.5--2.0 keV.
\end{table*}

\subsection{Correlation analysis}

From the summary of emission components in Table~\ref{RGS:PE}, we find there seem to be two distinct emitters. One emitter has a velocity of a few thousand km s$^{-1}$ but is only found in 2000-2009. The other, with a velocity of a few hundred km s$^{-1}$, seems always be there after 2006. The fast one may be associated with the wind, given the similarity in velocity, and the other is likely associated with BLR/NLR material as found by~\citet{Longinotti2008}.

\begin{figure}
    \centering
    \includegraphics[width=\linewidth]{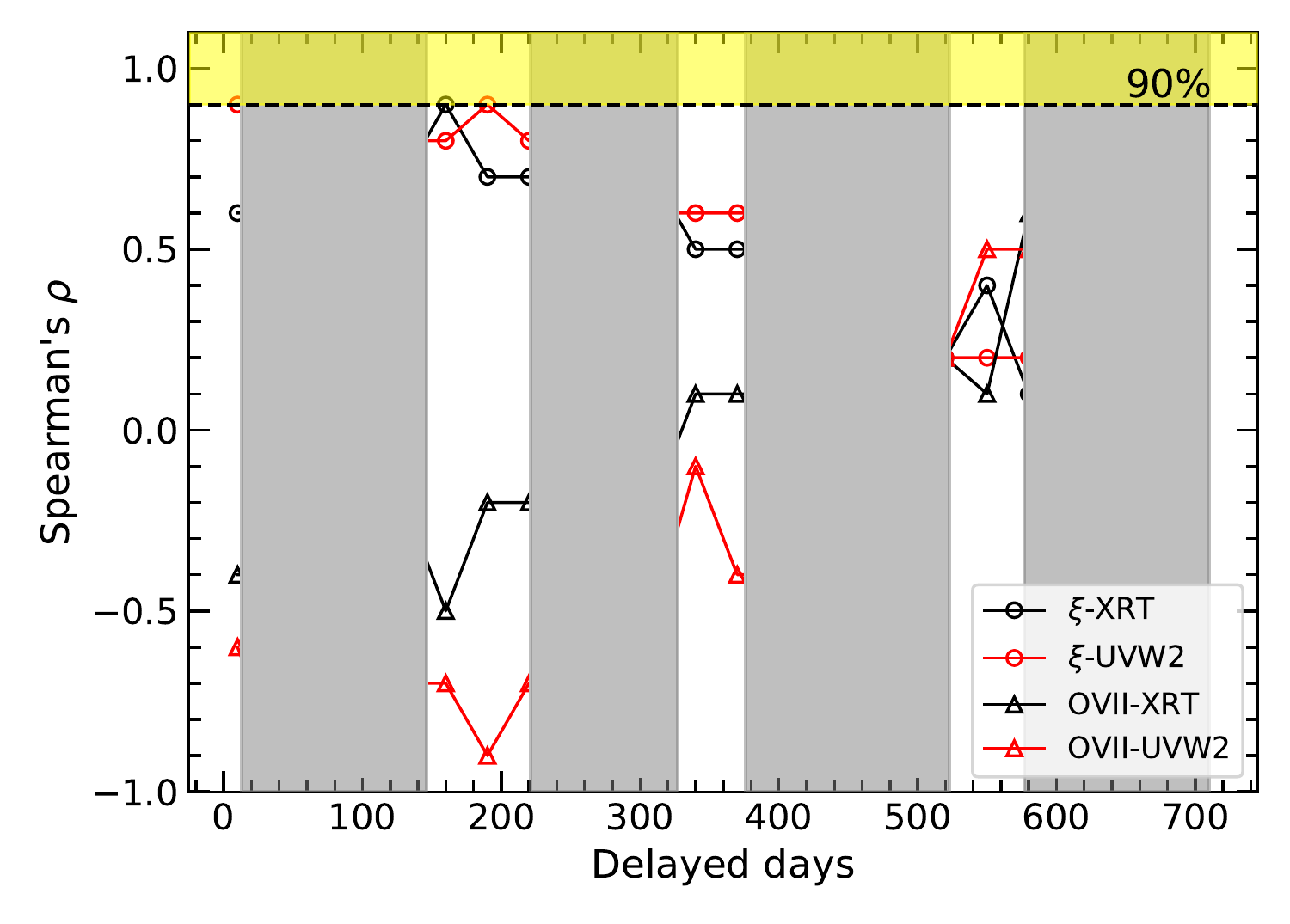}
    \caption{Spearman's $\rho$ between the source count rate observed by \textit{Swift} XRT/UVW2 and the ionization parameter of photoionized emission material as a function of time lag. The black circles represent the correlation coefficient between the ionization parameter and XRT count rate and red circles are for the correlation between ionization parameter and UVW2 magnitude. The results from tracing the \ion{O}{vii} f line are denoted by triangles. The grey regions mark intervals in which the correlation analysis is unreliable, since the \textit{Swift} monitoring is interrupted for 3 months each year. The yellow shaded region above the horizontal dashed line marks where the null-hypothesis probability is less than 10\%.}
    \label{PE:correlation}
\end{figure}

One way to locate the emitter is to see if the gas properties are responding to the source brightness with some delay. The delayed time can be an indication of the distance between the emitter and the X-ray source. We try to look for such delayed response by investigating if the the ionization parameter of the emitting gas is correlated with the source luminosity some days (the time delay) before each observation. If using luminosity from the \textit{XMM-Newton} broadband spectra for the correlation analysis, we can not get any information on the time delay. So we use the data from the \textit{Swift} monitoring (see Figure~\ref{lightcurve}) to indicate the source luminosity. We obtain the source count rates (or magnitude) of Mrk~335 before each observation (since 2007) at a series of time scales (e.g. 7, 14 and 21 days before each observation) by interpolating the \textit{Swift} XRT/UVW2 monitoring. Then we calculate Spearman's rank correlation coefficient between this source count rates and the ionization parameter of the low velocity emitter for each timescale. The correlation coefficient ($\rho$) can be used to indicate the probability that the observed correlation is obtained by chance ($P$-value). The correlation coefficient as a function of delayed timescale are displayed in the Fig.~\ref{PE:correlation}. The grey regions in Fig.~\ref{PE:correlation} represent timescales with unreliable correlation coefficient due to the Sun constraint for \textit{Swift}. The unreliable intervals cover a large fraction of 0 to 700 days timescales and prevent us from telling if there is correlation between the source and the emitting gas.

Tracing the variability of emission lines (e.g. \ion{O}{vii} f) is also a way to measure the scale of emitting region~\citep[e.g.][]{Detmers2009}. We conduct the same correlation analysis as above between the line flux of \ion{O}{vii} forbidden and the source brightness. The line flux is obtained by modelling the \ion{O}{vii} forbidden line (22.1~\AA~at rest frame) with a zero width gaussian outside the warm absorber. We have also checked that the warm absorber did not significantly affect the \ion{O}{vii} forbidden line. The results are also shown in Fig.~\ref{PE:correlation}. There is no surprise that the unreliable regions still make it hard to determine whether the correlation exists.

\section{discussion}


\subsection{The line emitting gas}

The current data does not permit a detailed tracing of the response of the emitting gas to the source variability. We can still compare the variation amplitude ($R=$maximum/minimum) of the emission lines and X-ray flux of the source. For the ten \textit{XMM-Newton} observations in the last 20 years, the variation amplitude of the X-ray flux (in 0.5--10 keV) of Mrk~335 is $R_{\rm X}=29.9$. However, considering the total flux of all \textsc{pion} components for each observation, the amplitude of the emission flux is $R_{\rm \textsc{pion}}=2.0$, which is significantly lower than that of the X-ray continuum. This conclusion holds even if only the flux of \ion{O}{vii} forbidden line is considered ($R_{\rm \ion{o}{vii}~f}=1.8$). This is similar to what have been found by~\cite{Komossa2020}. In addition to the lack of correlation between X-ray and UV during the 2020 flare of Mrk 335, they also find that the variation amplitude of the reverberation-mapped \ion{He}{ii} of Mrk~335 is smaller than that in X-rays. \cite{Komossa2020} explained this with an absorption scenario, in which the observed X-ray is different from what is seen by the material out of the line of sight. Variability in the observed X-ray continuum driven by absorption is commonly seen in AGNs~\citep[e.g.][]{Risaliti2011, Walton2014} and varying absorption has been clearly detected in Mrk 335~\citep{Longinotti2013}. If the X-ray continuum is altered by cold absorption material along the line of sight, but the line emitting gas is seeing the unaffected (or affected by different absorption material) continuum, there could be discordance between the observed X-ray luminosity and the X-ray lines. The other possibility is that the emission lines are from an extended (perhaps low density) gas. The variation amplitude will be reduced due to the delayed response of different parts of the gas to the variations of the X-ray continuum~\citep{Peterson1993}.

\begin{table}
\centering
\caption{Best-fit values for Athena simulations of Mrk~335}
\label{simulation}
\renewcommand\arraystretch{1.5}
\begin{tabular}{cccc}

\hline\hline
Component  & Parameter & 2009 & 2019 Dec  \\
\hline
WA (\romannumeral1) & $N_{\rm H}$ (cm$^{-2}$) &  $1.84_{-0.09}^{+0.3} \times 10^{22}$ & $7.1_{-0.3}^{+0.3} \times 10^{20}$  \\
   & $\log(\xi)$ &  $3.156_{-0.007}^{+0.04}$   &  $1.369_{-0.006}^{+0.006}$  \\
   & $v_{\rm out}$ (km s$^{-1}$) & $-5220_{-20}^{+20}$  &  $-5800_{-10}^{+10}$  \\
\hline
WA (\romannumeral2) & $N_{\rm H}$ (cm$^{-2}$) & $5.95_{-0.12}^{+0.11} \times 10^{21}$  &  $-$ \\
   &  $\log(\xi)$ &  $2.359_{-0.003}^{+0.006}$  &   $-$  \\
   &  $v_{\rm out}$ (km s$^{-1}$) & $-5200_{-10}^{+10}$  &  $-$  \\
\hline
WA (\romannumeral3) & $N_{\rm H}$ (cm$^{-2}$) & $2.07_{-0.04}^{+0.04} \times 10^{21}$  & $-$ \\
   &  $\log(\xi)$ &  $1.320_{-0.006}^{+0.004}$  &  $-$  \\
   &  $v_{\rm out}$ (km s$^{-1}$) & $-5200_{-10}^{+10}$  &  $-$  \\
\hline
PE (\romannumeral1) & $\log(\xi)$ &  $1.350_{-0.015}^{+0.012}$  &  $1.025_{-0.014}^{+0.014}$  \\
   & Density (cm$^{-3}$)  &  $5_{-4}^{+6} \times 10^{9}$  &  $1.00_{-0.17}^{+0.16} \times 10^{10}$ \\
   &  $v_{\rm out}$ (km s$^{-1}$) &  $-6000_{-10}^{+10}$ & $-150_{-5}^{+5}$  \\
   &  Flux (10$^{-13}$ erg s$^{-1}$ cm$^{-2}$) &  $1.69_{-0.06}^{+0.06}$  &   $1.53_{-0.02}^{+0.03}$   \\
\hline  
PE (\romannumeral2)   & $\log(\xi)$  &  $1.294_{-0.009}^{+0.009}$  &   $2.174_{-0.007}^{+0.007}$  \\
   &  Density (cm$^{-3}$) &  $4_{-3}^{+9} \times 10^{9}$  &  $6_{-3}^{+6} \times 10^{9}$  \\
   &  $v_{\rm out}$ (km s$^{-1}$) &  $-850_{-15}^{+15}$ &  $-200_{-5}^{+5}$   \\
   &  Flux (10$^{-13}$ erg s$^{-1}$ cm$^{-2}$) &  $1.02_{-0.02}^{+0.02}$  &  $1.370_{-0.02}^{+0.02}$  \\
\hline  
PE (\romannumeral3)   &  $\log(\xi)$ &  $1.196_{-0.011}^{+0.011}$  &  $-$  \\
   & Density (cm$^{-3}$)  &  $9_{-7}^{+5} \times 10^{9}$  &  $-$  \\
   &  $v_{\rm out}$ (km s$^{-1}$) &  $400_{-160}^{+120}$  &  $-$  \\
   & Flux (10$^{-13}$ erg s$^{-1}$ cm$^{-2}$) &  $1.38_{-0.03}^{+0.03}$  &  $-$  \\
\hline  
& C-stat/Dof & 4165/4207 & 4088/4215  \\
\hline  

\end{tabular}
\end{table}

\subsection{The warm absorbers}
The warm absorber outflows in AGNs can be produced by radiation pressure, magnetic force or thermal pressure. Both relativistic reflection and partial covering absorption models have been applied to Mrk~335 since its flux drop in 2007~\citep{ONeill2007, Grupe2007}, but it is only in the 2009 RGS spectrum that clear narrow absorption lines are detected, and the existence of WA in Mrk~335 has since been well established~\citep{Longinotti2013, Longinotti2019, Parker2019}. The best-fit parameters for the warm absorbers of Mrk~335 are shown in Table~\ref{RGS:WA}. All warm absorbers, except the one from the 2000 observation, exhibit velocities larger than 5000 km s$^{-1}$. Since a thermal wind can only be launched at large radii ($R \geq 10^5 R_{\rm g}$) and reach a maximum velocity $\sim$ 1000 km s$^{-1}$~\citep{Dorodnitsyn2008, Mizumoto2019}, it is less likely the driving force for the warm absorber outflows in Mrk~335.

By applying the three warm absorbers model from the 2009 spectrum to previous observations and assuming photoionization equilibrium of the gas, \citet{Longinotti2013} find no correlation between the warm absorber's properties and the flux variability of Mrk~335. We calculate the Spearman's rank correlation coefficient ($\rho$) between the source fluxes (0.5--10 keV, Galactic absorption corrected) and ionization parameters of warm absorbers. The source fluxes are obtained by fitting the broadband EPIC pn spectra (without partial covering). We find a high probability of a correlation coefficient (p=40\%) being obtained by chance from uncorrelated data, which confirms the finding by~\citet{Longinotti2013}. Note that we use the second layer of 2009 warm absorbers to conduct the correlation analysis, but we have checked that the result is not affected by other choices. 

We also test the possibility that the intrinsic luminosity of Mrk~335 is altered by a partial covering absorber. This can lead to underestimation of the source luminosity if the partial covering absorption is not corrected. We fit each EPIC pn spectrum with one partial covering component and get the intrinsic X-ray luminosity (correcting the partial covering absorption). However, there is still no correlation (p>25\%) between this luminosity and WA ionization state.

In the 2009 spectrum where the warm absorption feature is most prominent, 3 warm absorbers are needed to fit the data. However, only one layer of warm absorber is needed for other data in the low and high flux state. It is possible that during these states, the WAs are still there, but being de-ionized (or overionized), showing no detectable features over the continuum. We test this scenario by fitting the 3 WAs in 2009 spectrum to other datasets, keeping the column density and outflowing velocities at the best-fit values in 2009 and allowing only the ionization parameter to vary. The result shows that the two highest ionization components are not needed by the fit, since their ionization parameters are pegged at the upper limit ($\log(\xi) = 5$ for our \textsc{xstar} model) and have minor detectable effect given the column density. It could be due to that the two components are out of the line of site due to their transverse motion, as suggested by~\cite{Longinotti2013}. In addition, the 3 WAs found in 2009 spectrum share a consistent velocity but have very different ionization states. This pattern is similar to the WAs in the NLS1 I~Zwicky~1~\citep{Silva2018}, which are interpreted as clumpy clouds with the more ionized skin layer facing the source. The variation of the column density is not a surprise if some layers of the WAs in Mrk~335 also have clumpy structure.

The lowest ionization component, however, is formally required by the 4 datasets in 2015 and 2018--2019 low flux state, with the ionization parameter ($\log(\xi)$) to be $1.39_{-0.16}^{+0.19}$, $0.49_{-0.08}^{+0.10}$, $0.45_{-0.10}^{+0.10}$ and $0.57_{-0.12}^{+0.09}$ respectively. Note that the emission lines (especially the O VII triplets) are not well fitted in this test, even though the statistics are acceptable. From 2009 to 2015, the observed X-ray luminosity of Mrk~335 decreases by half. The ionization parameter of the WA should decrease if we are seeing the same material in photoionization equilibrium. Nevertheless, from this test, we can not tell if the ionization parameter decreases in 2015 (it changes from $1.32_{-0.07}^{+0.03}$ to $1.39_{-0.16}^{+0.19}$), while the WA might have changed ionization in 2018. If we assume that the same WA is seen in the 4 datasets in 2015 and 2018--2019 low state and its ionization state responses to the flux drop in 2018, then the recombination timescale of the WA must be shorter than 31 months. Adopting equation (2) in~\cite{Reynolds1995} and assuming electron temperature $T=10^5$ K would give the lower limit of the electron density: $n>6\times10^3$ cm$^{-3}$ (consider oxygen). This in turns gives the upper limit of the location of the WA: $r<8.4$ pc using the definition of the ionization parameter. This range is consistent with previous studies~\citep[e.g.][]{Longinotti2013,Longinotti2019} but the large value of the upper limit does not help to determine the origin of the WA. To study the WA in Mrk~335 through its variations, we would need more observations in the future to probe both short and long timescale variations.

\subsection{The broadband spectra}
The fitting of broad band spectra of Mrk~335 often needs to include a distant reflection component~\citep[e.g.][]{Parker2014, Wilkins2015, Gallo2019}. One of the prominent signatures of distant reflection is the narrow core of iron K$\alpha$ peaked around 6.4 keV, which is almost ubiquitously detected in X-ray spectra of AGNs~\citep{Shu2010, Singh2011}. The line could be produced in the molecular torus~\citep[e.g.][]{Nandra2006, Shu2011}, the outer part of the disk~\citep[e.g.][]{Petrucci2002} or even the broad-line region~\citep[e.g.][]{Bianchi2008}. From the latest \textit{XMM-Newton} observation of Mrk~335, we find the distant reflection component still dominates the spectrum above 5 keV, although the source flux drops dramatically in July 2018. It suggests that the reprocessing material might be at least 17 light-months (0.43~pc) away from the black hole. 

The photoionized emission lines still dominates the soft energy band in 2019 December. This might indicate that the distance between the emission material and X-ray source is also larger than 17 light-months. If this is the case, the density of the hot component has to be lower than $\sim 2.5\times 10^5 {\rm cm}^{-3}$. It is possible that the emission lines come from large scale low density gas such as the torus atmosphere. However, we also note that the possible existence of partial covering absorption in the low state~\citep{Parker2019} makes it more difficult to constrain the scale of the line emitting gas. The partial covering material absorbs mostly on the low energy part of the observed spectrum (see Figure 10 of~\cite{Parker2019}). The line emitting gas, however, could be seeing a less affected SED and thus produce lines dominating the soft X-ray band. There are several pieces of evidences that support the existence of partial covering absorber in Mrk~335. First, we do not see long-term correlation between the observed X-ray luminosity and the X-ray line emitting gas. The 2020 brightening of Mrk~335 can also be explained as variations of partial covering absorption~\citep{Komossa2020}. 

Note that we do not expect absorption to be the only reason for the observed variability of Mrk~335, since there is also evidence of change of corona geometry~\citep[e.g.][]{Wilkins2015} and correlation between X-ray and UV for some segments of the light curve~\citep{Gallo2018}, suggesting intrinsic source flux change. We have also seen an iron K lag~\citep{Kara2013} and a soft excess dip in the variance spectrum~\citep{Igo2020} of Mrk~335, which are hard to explain if the variability is dominated by absorption. In addition, we never see a an absorption K edge in the spectra of Mrk~335. So what we are seeing could be occasional obscuration events that, to some extent, play a role in the long-term variability of Mrk~335. Given the fact that we have only 5 data sets to test the correlation between the observed X-rays and the emitting gas, some of these observations being affected by absorption could break the possible correlation. More deep observations in the future and flux-resolved correlation analysis will certainly help to better understand the photoionized emitting gas in Mrk~335.

\subsection{Simulation with \textit{Athena}}
\begin{figure}
    \centering
    \includegraphics[width=\linewidth]{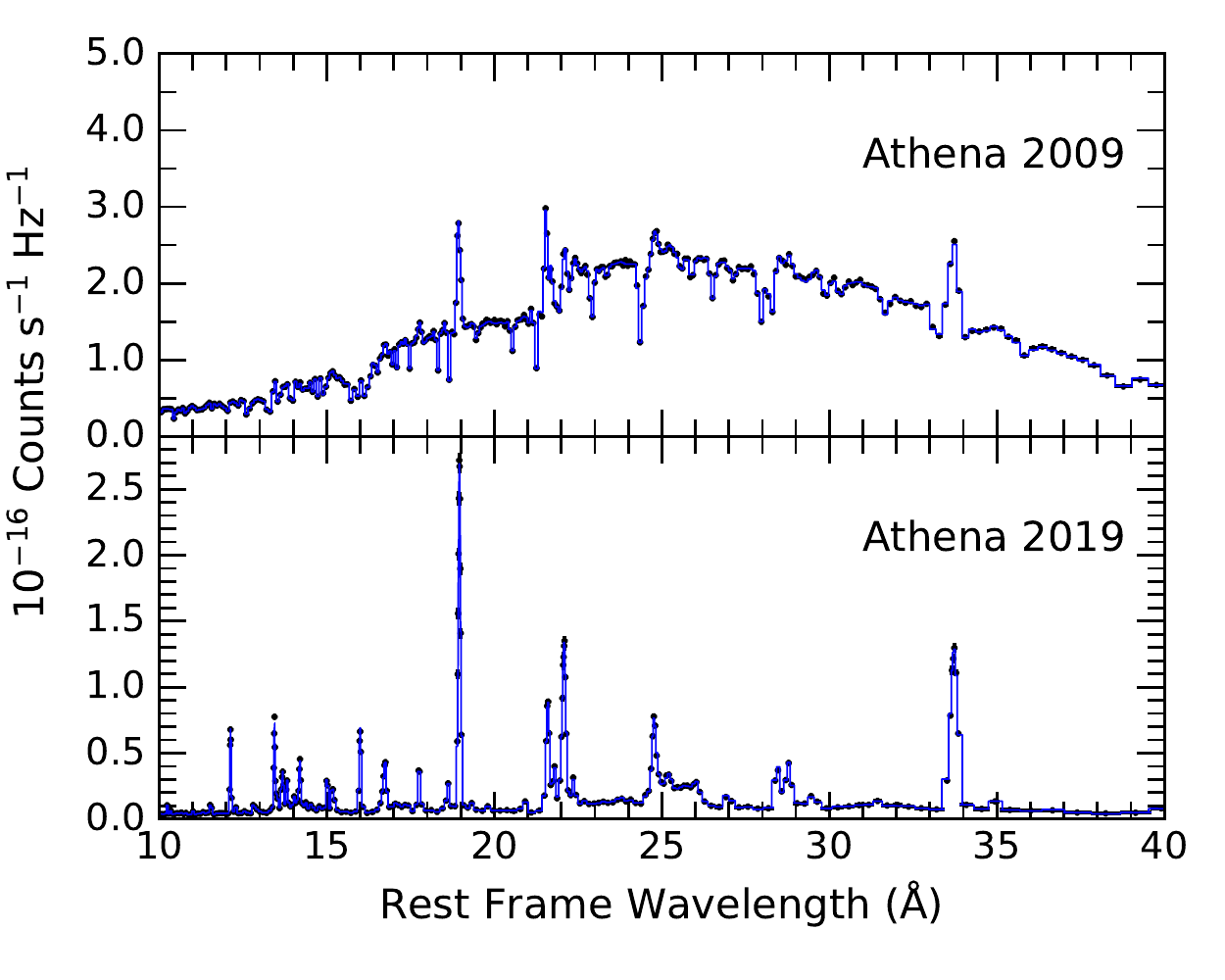}
    \caption{Simulation of the 2009 (upper) and 2019 (lower) spectra of Mrk~335. The black dots represent the simulated spectra and blue lines for best-fit model.}
    \label{simu}
\end{figure}

To see if future missions like \textit{Athena}~\citep{Nandra2013} can better reveal properties of the emitting and absorption material, we simulate absorption dominated (2009) and emission dominated (2019 Dec) spectra of Mrk~335 using response files of \textit{Athena} X-ray Integral Field Unit (X-IFU). Since density of the photoionized gas is not well constrained in our study with RGS data, we fixed it at $1\times 10^{10}$ cm$^{-3}$ for the simulation, and other parameters are set to their best-fit values. Spectra with exposure of 100 ks are simulated and fitted using C-stat. Constraints of parameters are shown in Table~\ref{simulation} and simulated spectra are shown in Figure~\ref{simu}. We find that, in both cases, Athena can better constrain the density and ionization parameter of photoionized emitters (with much lower error bars). The gas density can already indicate if the emission is from the BLR or NLR~\citep{Netzer1990}. Moreover, the density and ionization parameter can be used to measure the distance between the emitting gas and the X-ray source using the definition of ionization parameter ($\xi=L/nr^2$), if the flux history of the source is known. For instance, considering the first emission component in the last column of Table~\ref{simulation}, if the source is not variable (so no need to consider delay by light traveling and recombination), the X-ray emission lines would come from 0.01 pc away from the central black hole (assuming source luminosity 10$^{44}$ erg s$^{-1}$) with the error bar around 10\%.

\section{conclusion}

We have analyzed all available RGS spectra and broad band spectra of the latest \textit{XMM-Newton} observation of the NLS1 Mrk~335. The purpose is to study the behavior of line emitting material and the warm absorbers of this source. Our main findings are:

\begin{itemize}
\item With current data, we are not able to trace the response of the line emitting gas to the X-ray flux of Mrk~335. However, the emission lines are significantly less variable than the X-ray continuum. It could be due to the observed X-ray continuum is affected by absorption or the line emitting gas has an extended geometry.
\item We confirm the finding of~\citet{Longinotti2013} that the warm absorber in Mrk~335 is not responding to the source variability.
\item The dominance of distant reflection above 5 keV in 2019 December suggests that these components have not responded to the source flux drop in 2018 July, which indicates that the torus might be at least 17 light-months (0.43~pc) away from the X-ray source. 
\item The emission lines in the soft band of the 2019 December spectrum also have not responded to the flux drop. But it is more difficult to constrain the scale of the emitting gas using this lack of response, given the possible existence of partial covering absorber.
\item Future X-ray missions like \textit{Athena} are able to better determine the ionization state and density of the line emitting gas of Mrk~335, both in absorption dominated or emission dominated cases.
\end{itemize}


\section*{Acknowledgements}

The work of H.L. and C.B. was supported by the Innovation Program of the Shanghai Municipal Education Commission, Grant No.~2019-01-07-00-07-E00035, the National Natural Science Foundation of China (NSFC), Grant No.~11973019, and Fudan University, Grant No.~JIH1512604. J.J. acknowledges the support of the Tsinghua Shuimu Program and the Tsinghua Astrophysics Outstanding Fellowship. This work made use of data supplied by the UK Swift Science Data Centre at the University of Leicester and data from \textit{XMM-Newton}, an ESA science mission with instruments and contributions directly funded by ESA Member States and NASA.

\section*{Data Availability}

All the data can be downloaded from HEASARC.




\bibliographystyle{mnras}
\bibliography{bibliography} 






\bsp	
\label{lastpage}
\end{document}